\documentclass[conference]{IEEEtran}

\IEEEoverridecommandlockouts
\def\BibTeX{{\rm B\kern-.05em{\sc i\kern-.025em b}\kern-.08em
		T\kern-.1667em\lower.7ex\hbox{E}\kern-.125emX}}

\usepackage{dblfloatfix}    

\usepackage{cite}
\usepackage{amsmath,amssymb,amsfonts}
\usepackage{bbm}
\usepackage{epsfig}
\usepackage{graphicx}
\usepackage{textcomp}
\usepackage{xcolor}
\usepackage{colortbl} 
\usepackage{url} 
\usepackage[top=0.75in, bottom=1.08in, left=0.635in, right=0.635in]{geometry}

\usepackage{bm}
\usepackage[normalem]{ulem} 
\usepackage{supertabular}
\usepackage{amsthm}
\theoremstyle{definition}

\usepackage[linesnumbered,ruled,vlined]{algorithm2e}
\usepackage{algorithmicx} 
\usepackage{pbox}
\usepackage{float} 
\usepackage{subcaption} 
\DeclareMathOperator{\EX}{\mathbb{E}}
\newcommand\numberthis{\addtocounter{equation}{1}\tag{\theequation}}

\usepackage{boldline,multirow}
\usepackage{array}

\usepackage{pbox}

\usepackage{dblfloatfix}

\usepackage{pdfpages}

\begin{document}

\title{Channel Estimation with Asynchronous Reception for User-Centric Cell-Free MIMO Systems}
%
%
\author{Xuyang Sun\IEEEauthorrefmark{1}, Hussein Ammar\IEEEauthorrefmark{2}, Raviraj Adve\IEEEauthorrefmark{1}, Israfil Bahceci\IEEEauthorrefmark{3}, Gary Boudreau\IEEEauthorrefmark{3}
	\\
	\IEEEauthorrefmark{1}Dept.~of Elec.~and Comp.~Eng., University of Toronto, Toronto, ON, Canada\\
	
	\IEEEauthorrefmark{2}Dept.~of Elec.~and Comp.~Eng., Royal Military College, Kingston, ON, Canada\\
	\IEEEauthorrefmark{3}Ericsson Canada, Ottawa, ON, Canada \\
	Email: marcoxuyang.sun@mail.utoronto.ca; hussein.ammar@rmc-cmr.ca; rsadve@ece.utoronto.ca; \\
	 israfil.bahceci@ericsson.com; gary.boudreau@ericsson.com
	 \vspace{-2em}
}

\maketitle 


\begin{abstract}
The user-centric, cell-free wireless network is a promising next-generation communication system, but signal synchronization issues arise due to distributed access points and lack of cellular structure. We propose a novel method to recover synchronous pilot reception by introducing new pilot sequences and a matched filter window, enabling orthogonality even with asynchronous reception. Our approach mimics synchronous transmission by extending training sequences. Analysis shows asynchronous reception’s impact on channel estimation, and our method significantly improves performance with a small increase of training time overhead. Results demonstrate a 7.26 dB reduction in normalized mean square error and 40\% increase in data rate, achieving performance levels comparable to the synchronous case.
\end{abstract}

%

%
%
%
\section{Introduction}
\begin{figure}[t]
	\centering
	\includegraphics[width=1\columnwidth]{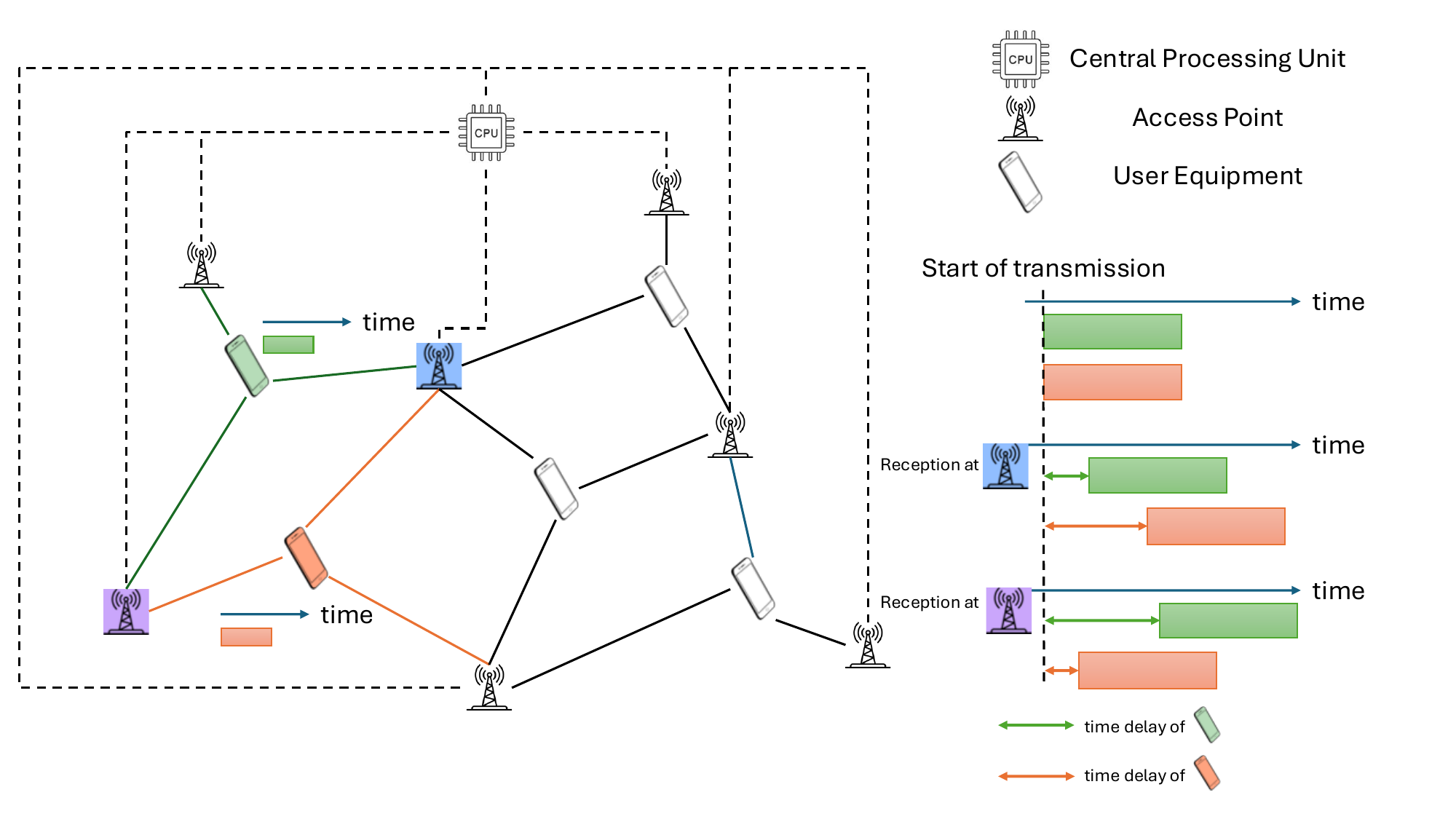}
    \vspace{-1em}
	\caption{Illustrating a cell-free system}
	\label{fig: Cell-Free System}
    \vspace{-1.5em}
\end{figure}

The user-centric cell-free wireless network architecture has been proposed for next generation of wireless communications beyond 5G~\cite{9349624,9113273}. This cell-free network offers significant advantages in terms of uniform spatial service, supreme quality of service, and deployment feasibility~\cite{9650567}. Compared to a traditional cellular system with a macro base station serving one or more defined fixed cells, a cell-free network dynamically assigns multiple access points (AP) to serve user equipments (UE), creating virtual cells for each UE. Specifically, UEs connect to multiple nearby APs, reducing transmission distance and hence path loss, and ensuring at least one AP can provide reliable service. Fig.~\ref{fig: Cell-Free System} shows a typical architecture of the cell-free networks in which UEs connect to nearby APs whose actions are coordinated, by a central unit~(CU) through a fronthaul interface. 

The optimization and design of cell-free systems face several challenges~\cite{9650567}. One significant challenge that has received limited attention is achieving seamless synchronization across all APs serving multiple UEs~\cite{8845768}. Due to the different timing misalignments and path lengths between multiple APs and the UEs~\cite{8969384}, synchronous reception is hardly feasible. This differs from a traditional cellular system wherein UE transmission timing may be advanced/delayed to ensure time synchronization with the serving BS~\cite{Nasir_2016}. The problem is further complicated by the fact that different UEs connect to different, but not disjoint, groups of APs, removing the concept of cells. Fig.~\ref{fig: Cell-Free System} illustrates the issue of asynchronous reception for cell-free systems. The serving clusters of the green and orange UEs overlap, but any attempt to synchronize reception at one serving AP could cause asynchronous reception at another. 

There have been limited work on the issue of synchronization in cell-free systems. In~\cite{8676341}, the authors provide an analysis of the lower bound of the downlink capacity with asynchronous reception. The authors in~\cite{10459246} develop an algorithm to cluster the APs into partially coherent clusters to achieve near-synchronous transmission within each cluster. The authors in~\cite{ganesan2023beamsyncovertheairsynchronizationdistributed} propose a novel protocol, over-the-air, transmitters could synchronize with each other without the help from a CU. These papers focus on downlink transmission mainly. In~\cite {9531352}, the authors analyze the impact of asynchronous reception on channel estimation and downlink transmission. However, the work does not provide any improvement methods. The authors note that asynchronous reception in the time \& phase domain is more difficult in cell-free systems, compared to cellular systems. One reason is that cooperation among the APs cannot be perfect due to the limited fronthaul capacity~\cite{10517887}.

Accurate channel estimation is crucial for performance. Typically, for multiuser channel estimation is the orthogonal pilot sequence assignment~\cite[Sec. 3.1.1]{8277240}. With proper synchronization, the orthogonality property eliminates inter-UE interference. Pilot contamination arises from co-pilot UEs who are usually kept spatially distant to minimize contamination. Under asynchronous reception, pilot sequences orthogonality is destroyed and interferes with the sequences of other UEs.

\begin{figure}[t] 
    \centering
    \begin{subfigure}[t]{0.85\columnwidth}
    \centering
    \includegraphics[width=1\columnwidth]{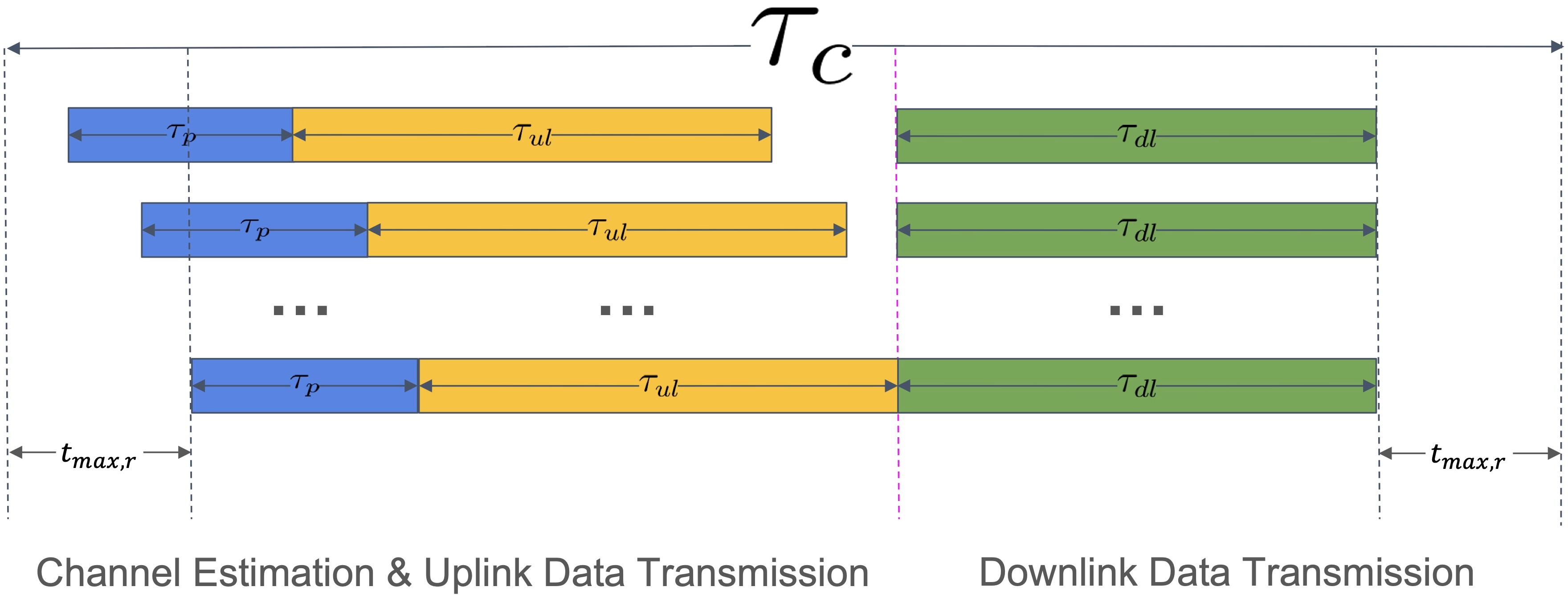}
    \caption{Asynchronous reception without guard times}
    \label{subfig: no guard time no ex}
    \vspace{0.3em}
    \end{subfigure}
    \begin{subfigure}[t]{0.85\columnwidth}
    \centering
    \includegraphics[width=1\columnwidth]{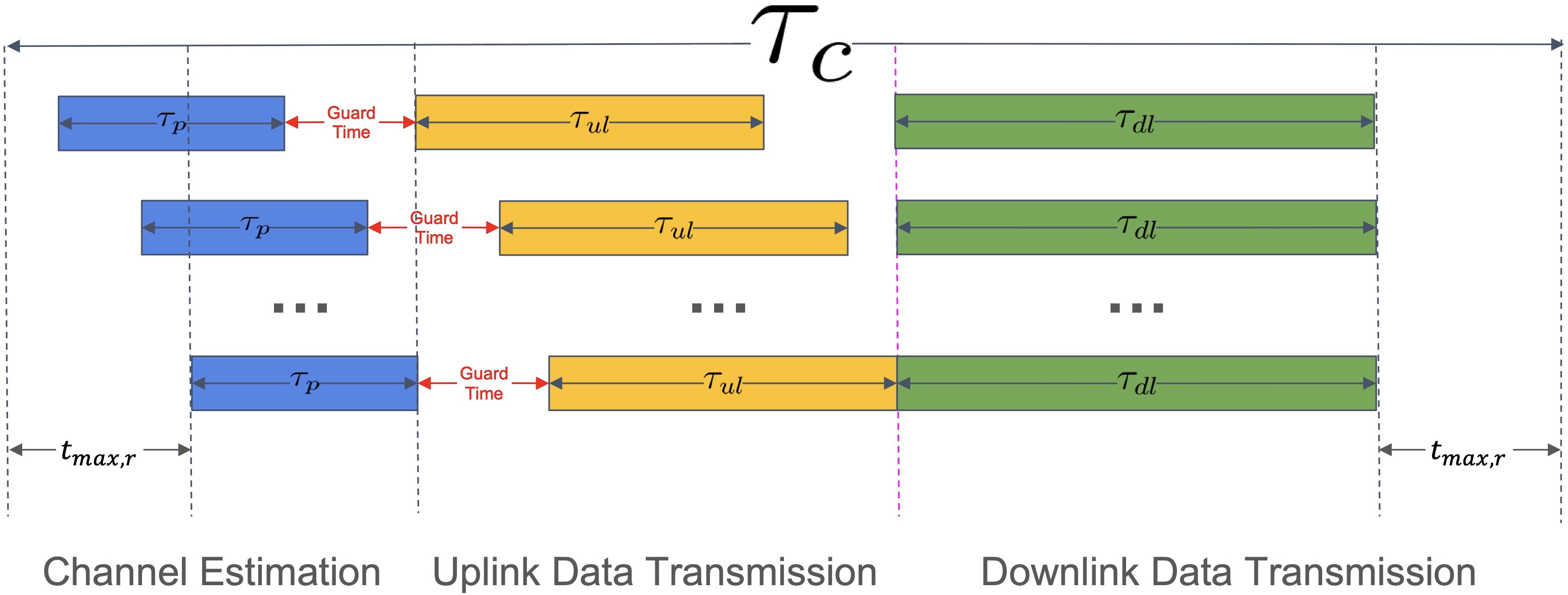}
    \caption{Asynchronous reception with guard time}
    \label{subfig: guard time no ex}
    \end{subfigure}
    \caption{Asynchronous reception  in one coherence block}
    \label{fig: Asynchronous Reception}
    \vspace{-1.8em}
\end{figure}

Fig.~\ref{fig: Asynchronous Reception} illustrates the asynchronous reception phenomenon. Each row in the figure corresponds to a UE transmitting to a chosen AP. In the figure, $\tau_{\rm p}$, $\tau_{\rm ul}, \tau_{\rm dl}$ and $\tau_{\rm c}$ refer to the time, measured in data samples, for pilot training, uplink data, downlink data, and coherence block, respectively. As seen in Fig.~\ref{subfig: no guard time no ex}, without a guard time the pilot transmissions of some UEs and the data transmission of others interfere with each other. Depending on the spatial distribution of these UEs, interference levels can surpass the desired signal levels, significantly impacting channel estimation and/or data reception. One possible simple solution is shown in Fig.~\ref{subfig: guard time no ex}, where a guard time eliminates the interference of uplink transmissions on the channel estimation. Importantly, however, this solution does not restore the orthogonality of pilot sequences.

This paper focuses on the asynchronous time and phase reception in the channel estimation phase. We investigate three different pilot schemes for channel estimation: random sequences, discrete Fourier Transform (DFT) sequences, and extended DFT sequences. We show that using extended DFT sequences achieves performance similar to synchronous transmissions. We also investigate two different cases: (i) uplink-pilot guard time (UPG): employs guard time between the channel estimation phase and the uplink transmission phase, and (ii) uplink-pilot no guard time (UPNG): there is no guard time between these phases. We compare the performance of these cases using the normalized-mean-squared error (NMSE).

The rest of the paper is organized as follows. Section~\ref{section:system model} introduces the system model. Section~\ref{section:channel_estimation} presents the channel estimation procedure for the three different pilot schemes and the mathematical analysis of their performance. Section~\ref{section:results} validates analysis with numerical results. Finally, Section~\ref{section:conclusion}, provides a conclusion of the paper and the lessons learned.

\textit{Notation:} italics to denote scalars, e.g., $r$, boldface to denote vectors, e.g., $\mathbf{h}$ and the calligraphic font to denote sets, e.g., $\mathcal{R}$. $\mathbb{C}^{M\times N}$ denotes the set of $M \times N$ matrices of complex numbers ($N=1$ denotes vectors). $\mathcal{CN}(\boldsymbol{\mu}, \mathbf{R})$ denotes the complex normal distribution with mean $\boldsymbol{\mu}$ and covariance matrix $\mathbf{R}$. $\mathbf{I}_M$ is the $M\times M$ identity matrix, and $\mathbb{E}[\cdot]$ is expectation.

\section{System Model}\label{section:system model}
\vspace{-0.5em}
We consider a user-centric cell-free system that operates in time-division duplex (TDD) mode. Our system follows a similar structure to Fig.~\ref{fig: Cell-Free System}. We use $\mathcal{R}$ and $\mathcal{U}$ to represent the sets of APs and UEs, where $|\mathcal{R}|$ and $|\mathcal{U}|$ are the number of APs and UEs, respectively. The APs and UEs are uniformly spatially distributed. Each AP is identified by its index $r$, $1 \leq r\leq |\mathcal{R}|$ while each UE by index $u$, $1 \leq u\leq |\mathcal{U}|$. Each AP is equipped with M antennas. Additionally, we establish a restricted area with a radius of $\gamma$~meters around each AP, preventing users from being located too close to the APs. This approach ensures reliable results by avoiding proximity effects and simulates a scenario where the APs are deployed at a height of $\gamma$~meters, serving users at ground level.
		

Each AP $r$ selects a set $\mathcal{U}_r \subseteq \mathcal{U}$ UEs to serve. These UEs can be chosen using a distance criterion or a threshold on the power loss (large-scale fading) between the AP and UE. We denote the set of APs serving UE $u$ as $\mathcal{R}_u$. We define the channel between AP $r$ and UE $u$ as $\boldsymbol{h}_{ru}\in\mathbb{C}^{M\times 1}$. The channel $\boldsymbol{h}_{ru} \triangleq \sqrt{\beta_{ru}\psi_{ru}}\boldsymbol{g}_{ru}$ where $\beta_{ru}$, $\psi_{ru}$, and $\boldsymbol{g}_{ru}\sim\mathcal{CN}(\mathbf{0},\boldsymbol{I}_M)$ accounts for the effects of path-loss, shadowing, and the Rayleigh fading component, respectively.

\vspace{-0.5em}
\section{Channel Estimation}\label{section:channel_estimation}
\vspace{-0.5em}
Channel estimation is performed within each coherence time block, where each UE~$u$ broadcasts a known training/pilot sequence $\boldsymbol{\phi}_u$. The choice of the pilot is a crucial contribution of this paper and will be discussed later. 
Channel estimation then proceeds through two steps: match-filtering (MF) and linear minimum mean squared error (LMMSE) estimation. In the synchronous case, UEs' signals arrive at APs at the same time and MF eliminates the interference from other UEs with orthogonal pilot sequences. The LMMSE step then minimizes the co-pilot interference. We focus on the asynchronous case where the UEs' transmissions arrive with different delays.

In Fig.~\ref{fig: Cell-Free System}, assuming UEs' transmissions start at time $0$\footnote{Any errors in transmission time due to differences in local clocks can be easily incorporated into the model.}, and the signal sequence received from the UE $u$ at the AP $r$ is,
\begin{align}
    \begin{split}
        \boldsymbol{x}_{u,r,\rm aug} = [\boldsymbol{0}^{1\times t_{u,r}},\boldsymbol{\phi}_u,\boldsymbol{s}_{u}]
    \end{split} \label{eq:augmented}
\end{align}
Here, $\boldsymbol{0}^{1\times t_{u,r}}$ is a row vector of $t_{u,r}$ zeros where $t_{u,r}$ represents the discretized time delay for the signal of UE $u$ when received at AP $r$. The array $\boldsymbol{s}_u\in\mathbb{C}^{1\times (t_{\rm max,r} - t_{u,r})}$ is part of the uplink data sequence of the user $u$ in UPNG, $t_{\rm max,r}$ is the maximum time delay among all UEs at the AP $r$. In the UPG, we replace $\boldsymbol{s}_u$ with $\boldsymbol{0}^{1\times (t_{\rm max,r} - t_{u,r})}$. This augmented signal sequence of the UE $u$, $\boldsymbol{x}_{u,\rm aug}$ is viewed from the receiver's (AP) point of view. The use of an augmented version allows us to consider the time delay in the signal model. The overall signal received at AP $r$ during pilot transmissions is,
\begin{align}\label{basic_signal_model}
    \begin{split}
        \boldsymbol{Y}_r = \sqrt{p^{\rm ul}}\sum_{u\in\mathcal{U}}\boldsymbol{h}_{ru}\boldsymbol{x}_{u,r,\rm aug} + \boldsymbol{Z}_r
    \end{split}
\end{align}
Here, $\boldsymbol{Y}_r \in \mathbb{C}^{M\times (\tau_{\rm p}+t_{\rm max,r})}$ is the received signal matrix at the AP $r$, $\boldsymbol{Z}_r\sim\mathcal{CN}(0,\sigma^2\boldsymbol{I}_M)$ is the additive white Gaussian noise (AWGN) at AP $r$, with independent entries, and $p^{\rm ul}$ is the power used for the pilot and uplink data sequence. 

\vspace{-0.5em}
\subsection{Matched Filtering and LMMSE Estimation}
To time align the pilot sequences within $\boldsymbol{x}_{u,r,\rm aug}$ from~\eqref{eq:augmented}, we define the zero-padded version, $\boldsymbol{\phi}_{u,r,\rm MF}\in\mathbb{C}^{1\times(\tau_{\rm p}+t_{\rm max,r})}$ that will be used at AP $r$ to estimate the channel of user $u$:
\begin{align}
    \boldsymbol{\phi}_{u,r,\rm MF} = [\boldsymbol{0}^{1\times t_{u,r}},\boldsymbol{\phi}_u,\boldsymbol{0}^{1\times (t_{\rm max,r} - t_{u,r})}] \label{eq:MFsequenceAugmented}
\end{align}
Using~\eqref{eq:MFsequenceAugmented}, matched filtering can be expressed as
\begin{align}\label{MF}
    \begin{split}
    &\Breve{\boldsymbol{y}}_{ru} = \frac{1}{\sqrt{p^{\rm ul}}}\boldsymbol{Y}_r\boldsymbol{\phi}_{u,r,\rm MF}^H\\[-5 pt]
    &= \boldsymbol{h}_{ru}\tau_{\rm p} + \sum_{u'\neq u}\boldsymbol{h}_{ru'}\boldsymbol{x}_{u',r,\rm aug}\boldsymbol{\phi}_{u,r,\rm MF}^H + \frac{\boldsymbol{Z}_{r}\boldsymbol{\phi}_{u,r,\rm MF}^H}{\sqrt{p^{\rm ul}}}
\end{split}
\end{align}
As shown in~\eqref{MF}, the desired signal term is $\boldsymbol{h}_{ru}\tau_{\rm p}$, the interference term is $\boldsymbol{\iota}_{ru} = \sum_{u'\neq u}\boldsymbol{h}_{ru'}\boldsymbol{x}_{u',\rm aug}\boldsymbol{\phi}_{u,\rm aug}^H$, and the noise term is $\boldsymbol{Z}_{r}\boldsymbol{\phi}_{u,\rm aug}^H$. Then, using LMMSE estimation, the channel estimate can be expressed as follows,
\begin{align}\label{LMMSE}
    \begin{split}
        \hat{\boldsymbol{h}}_{ru} &= \boldsymbol{\Sigma}_{\Breve{\boldsymbol{y}}_{ru}\boldsymbol{h}_{ru}}\boldsymbol{\Sigma}_{\Breve{\boldsymbol{y}}_{ru}}^{-1}\Breve{\boldsymbol{y}}_{ru},
    \end{split}
\end{align}
where $\boldsymbol{\Sigma}_{\Breve{\boldsymbol{y}}_{ru}}$ is the covariance matrix of the interference plus noise and $\boldsymbol{\Sigma}_{\Breve{\boldsymbol{y}}_{ru}\boldsymbol{h}_{ru}}$ is the cross-covariance between the true channel and the signal in~\eqref{MF}.

\vspace{-0.5em}
\subsection{Choice of Sequences}
\vspace{-0.3em}
We consider three different pilot sequences:
\begin{align}
	\begin{split}
		\boldsymbol{\phi}_u =
		\begin{cases}
			\boldsymbol{\phi}_{u,\rm ran}\in\mathbb{C}^{1\times\tau_{\rm p}}\\
			\boldsymbol{\phi}_{u,\rm DFT}\in\mathbb{C}^{1\times\tau_{\rm p}}\\
			\boldsymbol{\phi}_{u, \rm DFT_{ex}}\in\mathbb{C}^{1\times(\tau_{\rm p}+\tau_{\rm ex})}\\
		\end{cases}
	\end{split}
\end{align}
where $\tau_{\rm p}$ is the pilot length, and $\tau_{\rm ex}$ is the extended length which will be specified later. 

The first choice, $\boldsymbol{\phi}_{u,\rm ran}$, is a sequence of unit-magnitude symbols with each entry having an independent discretized random phase. This choice provides a baseline and mimics popular choices like Zadoff-Chu sequences. These sequences are, statistically, mutually orthogonal. The second choice, $\boldsymbol{\phi}_{u,\rm DFT}$, is chosen from the DFT matrix. The third one, which will be crucial in this paper, is a cyclically extended pilot sequence $\boldsymbol{\phi}_{u, \rm DFT}$ of length $\tau_{\rm p} + \tau_{\rm ex}$ samples.

We now analyze each of these choices of pilot sequences. 

\subsection{Random Sequences}
The first sequence is random, in which each element of the pilot sequence is unit power with a random phase defined as
\begin{align}
    \{\boldsymbol{\phi}_{u,\rm ran}\}_i = e^{j\theta_i}, \theta_i\sim U\{0,...,P-1\}\times\frac{2\pi}{P},
\end{align}
where the phase of each entry $i$, $\theta_i$, is independent and identically distributed. The power associated with each term of~\eqref{MF} can be expressed as:
\begin{align}
    \EX[(\boldsymbol{h}_{ru}\tau_{\rm p})^H(\boldsymbol{h}_{ru}\tau_{\rm p})] &= M\beta_{ru}\psi_{ru}\tau_{\rm p}^2\\[-2pt]
    \begin{split}
        \EX\left[\left(\boldsymbol{\iota}_{ru}\right)^H\left(\boldsymbol{\iota}_{ru}\right)\right] &= M\beta_{ru'}\psi_{ru'}\tau_{uu',r}
    \end{split}
    \\[-2pt]
    \EX[(\boldsymbol{Z}_{r}\boldsymbol{\phi}_{u,r,\rm MF}^H)^H(\boldsymbol{Z}_{r}\boldsymbol{\phi}_{u,r,\rm MF}^H)] &= \frac{M\sigma^2\tau_{\rm p}}{p^{\rm ul}}
\end{align}
Here, $\tau_{uu',r}$ is the overlap time between the sequences received from UE $u$ and UE $u'$ at the AP $r$. We see that as the pilot length $\tau_{\rm p}$ increases, the desired signal power increases quadratically, while the interference and noise power increase linearly, i.e., as expected, increasing the pilot length improves channel estimation. In~\eqref{LMMSE}, for the random sequence, $\boldsymbol{\Sigma}_{\Breve{\boldsymbol{y}}_{ru}\boldsymbol{h}_{ru}}$ and $\boldsymbol{\Sigma}_{\Breve{\boldsymbol{y}}_{rk}}$ can be expressed as,
\begin{align}
        \boldsymbol{\Sigma}_{\Breve{\boldsymbol{y}}_{ru}\boldsymbol{h}_{ru}} & = \tau_{\rm p}\beta_{ru}\psi_{ru}\boldsymbol{I}_M
        \\[-4pt]
        \boldsymbol{\Sigma}_{\Breve{\boldsymbol{y}}_{ru}} & = \beta_{ru}\psi_{ru}\tau_{\rm p}^2\boldsymbol{I}_M +  \sum_{u'\neq u} \beta_{ru'}\psi_{ru'}\tau_{uu',r}\boldsymbol{I}_M + \frac{\sigma^2\tau_{\rm p}}{p^{\rm ul}}\boldsymbol{I}_M
\end{align}

\subsubsection{UPG}
For the case employing a guard time, the sequence overlap time $\tau_{uu',r} = \tau_{\rm p} - |t_{u,r} - t_{u',r}|$.

\subsubsection{UPNG}
For the case without guard time, the sequence overlap time $\tau_{uu',r}$ is expressed as follows,
\vspace{-0.2em}
\begin{align}
\tau_{uu',r} =
    \begin{cases}
        \tau_{\rm p} - |t_{u,r} - t_{u',r}| &\text{$t_{u,r} \leq t_{u',r}$}\\
        \tau_{\rm p} &\text{$t_{u,r} > t_{u',r}$}
    \end{cases}
\end{align}

\begin{figure}[t] 
	\centering
	\includegraphics[width=0.85\columnwidth]{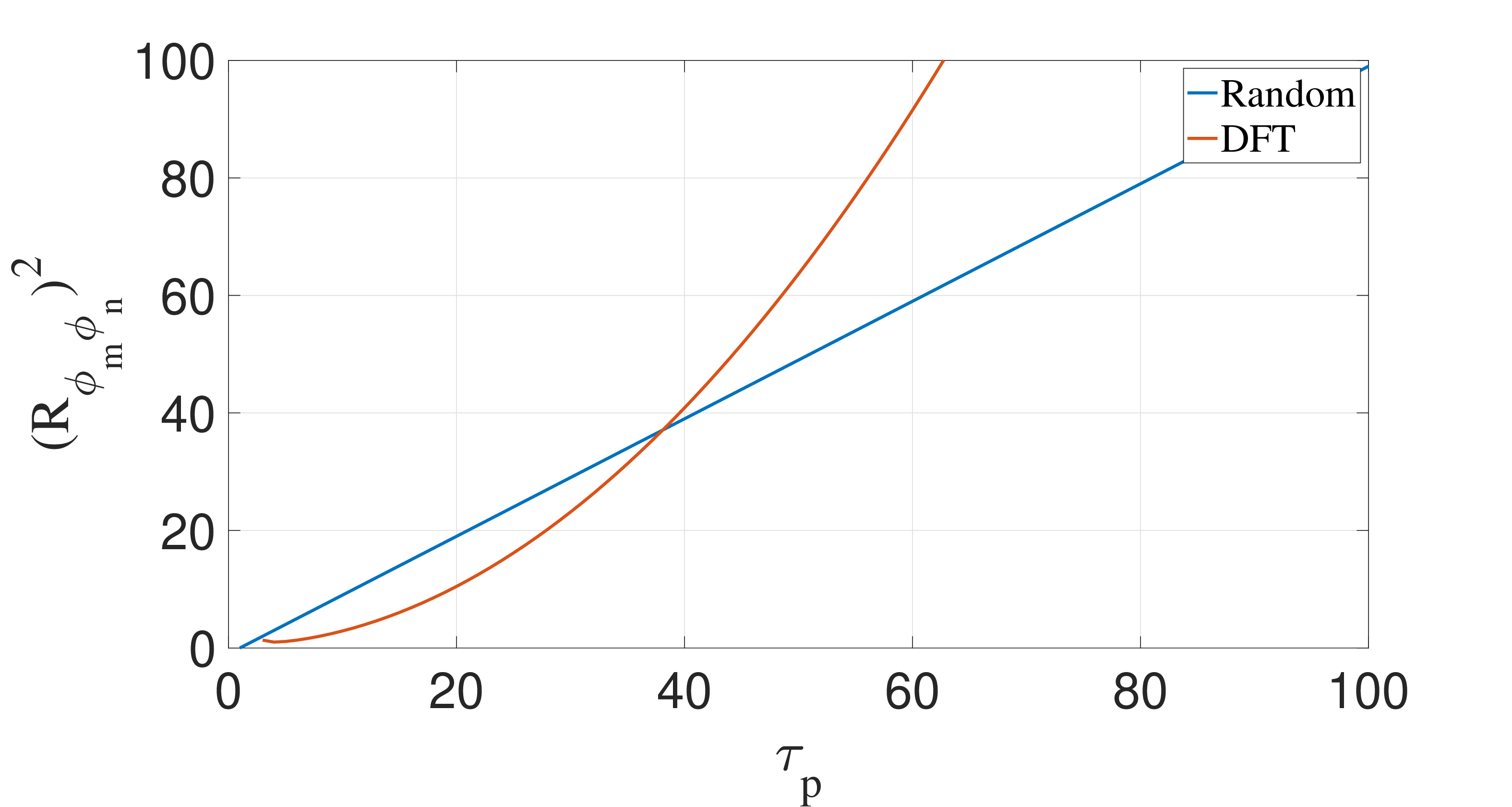}
	\vspace{-0.75em}
	\caption{Cross Correlation Comparison with a fixed time delay}
	\label{fig:Interference Comparison}
	\vspace{-1.5em}
\end{figure}

\vspace{-0.4em}
\subsection{DFT Sequences}
\vspace{-0.3em}
This sequence employs a DFT pilot sequence, chosen as a row of the DFT matrix, $\boldsymbol{\Phi}\in\mathbb{C}^{\tau_{\rm p}\times\tau_{\rm p}}$. The notation $[\boldsymbol{\Phi}]_m$ represents the $m^{th}$ row of $\boldsymbol{\Phi}$, i.e., $m$ is also the index of the pilot sequence. While the sequences associated with two different indices are mutually orthogonal, the orthogonality is destroyed if the signals are not perfectly synchronized.

In the synchronous case, $t_{u,r}$ is a constant among all UEs and APs and the MF eliminates interference from all other pilot sequences. In asynchronous case, due to the different time delay values $t_{u,r}$, interference cannot be eliminated since
\begin{align}
    \bigl[\boldsymbol{0}^{1\times t_{u,r}},[\boldsymbol{\Phi}]_m\bigl]\bigl[\boldsymbol{0}^{t_{u',r}\times1};[\boldsymbol{\Phi}]_n^H\bigl] \neq 0, \:\:\:\text{$t_{u,r}\neq t_{u',r}$}
\end{align}
\subsubsection{UPG}
Let us consider UE $u$ for which UE $u'$ is an interferer. The pilot sequences of UE $u$ and UE $u'$ are $\boldsymbol{\phi}_{u,\rm DFT} = [\boldsymbol{\Phi}]_m$, $\boldsymbol{\phi}_{u', \rm DFT} = [\boldsymbol{\Phi}]_n$ respectively. Then at AP $r$, the interference term after MF is, here $\omega = e^{-j2\pi/\tau_p}$.
\begin{align}\label{DFT cross}
    \boldsymbol{h}_{ru'}\boldsymbol{x}_{u',r,\rm aug}\boldsymbol{\phi}_{u,r,\rm MF}^H = \boldsymbol{h}_{ru'}\omega^{m(\tau_{\rm p} - \tau_{uu',r})}\frac{\omega^{(m-n)(\tau_{uu',r})} - 1}{\omega^{m-n}-1}
\end{align}
The power of the interference can be calculated as,
\begin{align}\label{DFT cross power}
    \begin{split}
        \EX\biggl[(\boldsymbol{h}_{ru'}\boldsymbol{x}_{u',r,\rm aug}&\boldsymbol{\phi}_{u,r,\rm MF}^H)^H(\boldsymbol{h}_{ru'}\boldsymbol{x}_{u',r,\rm aug}\boldsymbol{\phi}_{u,r,\rm MF}^H)\biggl]\\ = &M\beta_{ru'}\psi_{ru'}\left(R_{\boldsymbol{\phi}_u\boldsymbol{\phi}_{u'}}\right)^2\\
       \textrm{where~~} R_{\boldsymbol{\phi}_u\boldsymbol{\phi}_{u'}} = & \biggl(\frac{ \sin(\pi(m-n)\tau_{uu',r}/\tau_{\rm p})}{\sin(\pi(m-n)/\tau_{\rm p})}\biggl).
    \end{split}
\end{align}
The detailed derivation of~\eqref{DFT cross} and~\eqref{DFT cross power} is shown in the ``Appendix" section. For the interference power calculated above, the first three constants are the same as the ones with random sequences. The squared term is a quadratic function of $\tau_{\rm p}$, which implies that the interference increases quadratically as $\tau_{\rm p}$ increases. However, as we can see from Fig.~\ref{fig:Interference Comparison}, when the pilot length is short, e.g. $\tau_{\rm p} \leq 38$, DFT sequences generate less inter-user interference compared to the random sequences after an MF, using a short pilot length could save more time resources for data transmission. 

\begin{figure}[t] 
	\centering
	\begin{subfigure}[t]{0.4\linewidth}
		\centering
		\includegraphics[width=\textwidth]{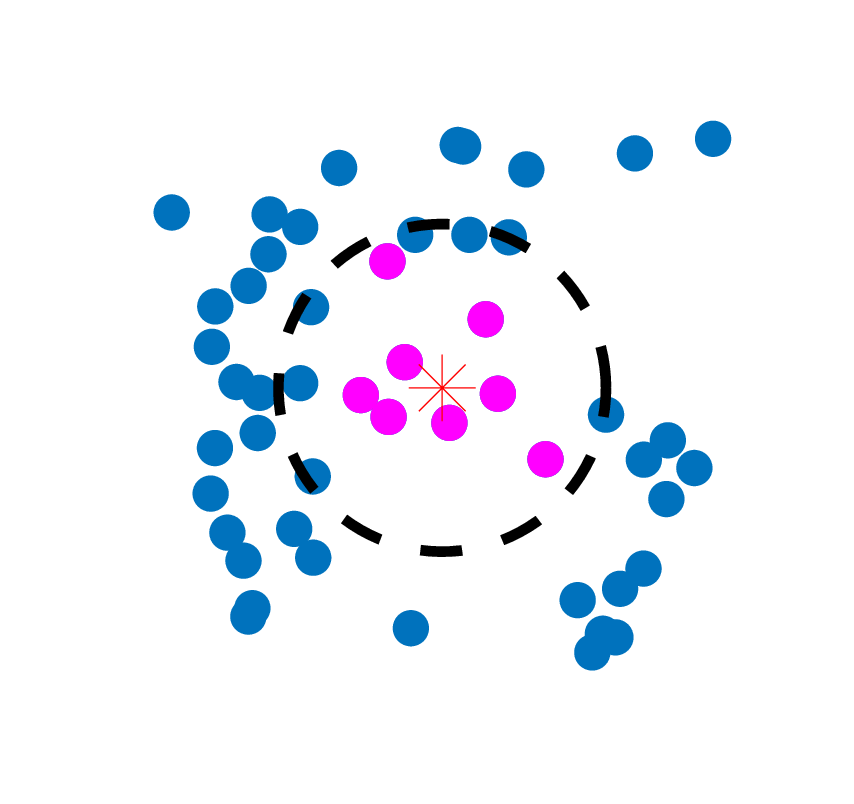}
		\vspace{-3em}
		\caption{$\mathcal{U}_r \subseteq \mathcal{S}_r$} \label{subfig:Significant User Region large}
	\end{subfigure}
	\begin{subfigure}[t]{0.4\linewidth}
		\centering
		\includegraphics[width=\textwidth]{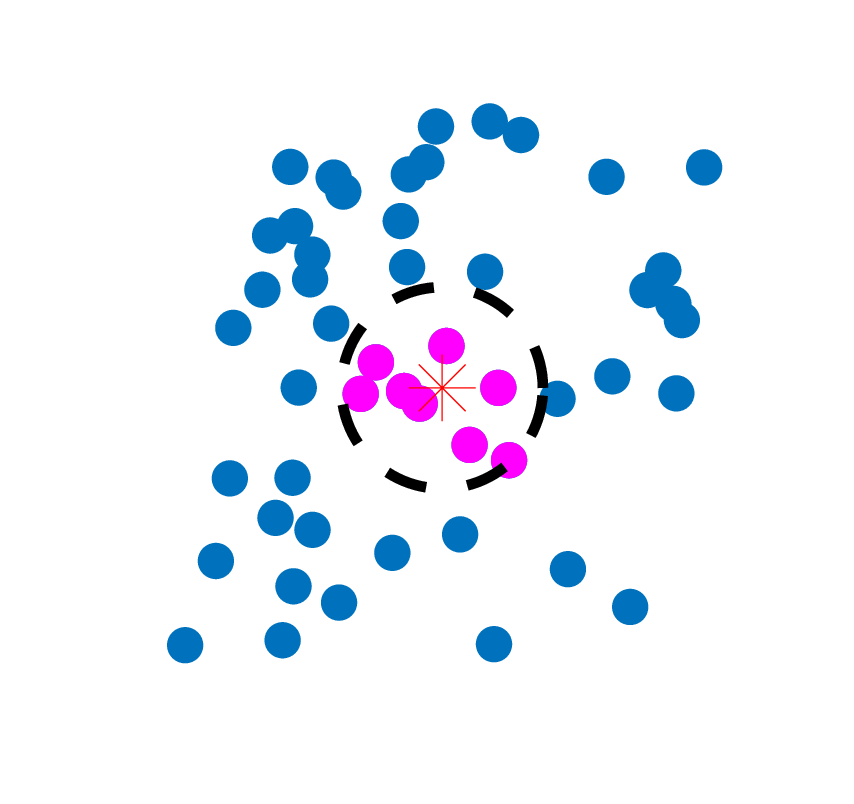}
		\vspace{-3em}
		\caption{$\mathcal{U}_r = \mathcal{S}_r$} \label{subfig:Significant User Region small}
	\end{subfigure}
	\caption{Significant User Region}
	\label{fig:Significant User Region}
	\vspace{-1em}
\end{figure}

\subsubsection{UPNG}
If there is no guard time between the channel estimation phase and the uplink transmission phase, then the interference term after MF at AP $r$ is
\begin{gather*}
        \hspace*{-2in}\boldsymbol{h}_{ru'}\boldsymbol{x}_{u',r,\rm aug}\boldsymbol{\phi}_{u,r,\rm MF}^H = 
        \\
        \begin{cases}
        \begin{aligned}
            \boldsymbol{h}_{ru'}\omega^{m(\tau_{\rm p} - \tau_{uu',r})}\frac{\omega^{(m-n)(\tau_{uu',r})} - 1}{\omega^{m-n}-1}
        \end{aligned}
        \hspace{0.2in} , \text{$t_{u,r} \leq t_{u',r}$}
            \\
        \begin{aligned}
            \boldsymbol{h}_{ru'}\omega^{m(\tau_{\rm p} - \tau_{uu',r})}\frac{\omega^{(m-n)(\tau_{uu',r})} - 1}{\omega^{m-n}-1} + \boldsymbol{h}_{ru'}\!\!\!
            \sum_{i = 0}^{\tau_{\rm p} - \tau_{uu',r} - 1}
            \!\!\! e^{j\eta_i}
        \end{aligned}
        \\ \hspace*{2in}  ,\text{$t_{u,r} > t_{u',r}$}
        \end{cases}
        \\
        \hspace{-0.2in}
        \text{where~~} 
        \eta_i \sim U\{0,...,{\rm lcm}(N_{\rm DFT}^m,N)\}\times\frac{2\pi}{{\rm lcm}(N_{\rm DFT}^m,N)}\\
        \hspace{0in}N_{\rm DFT}^m = \begin{cases}
            \begin{aligned}
        \frac{\tau_{\rm p}}{\mathrm{gcd}\left(m,\tau_{\rm p}\right)}
    \end{aligned}
    ~,~\text{$\frac{m}{\tau_{\rm p}}$ is rational}
    \\
    \begin{aligned}
        \tau_{\rm p}
    \end{aligned}
    \hspace{0.55in}~,~\text{$\frac{m}{\tau_{\rm p}}$ is irrational}
        \end{cases}\numberthis
\end{gather*}
Here, ${\rm lcm}(a,b)$ and ${\rm gcd}(a,b)$ represents the least common multiple and the greatest common divisor of $a$ and $b$, respectively. The power of the interference can be calculated as follows,
\allowdisplaybreaks
\begin{gather*}
    \hspace*{-0.5in}(\boldsymbol{h}_{ru'}\boldsymbol{x}_{u',r,\rm aug}\boldsymbol{\phi}_{u,r,\rm MF}^H)^H(\boldsymbol{h}_{ru'}\boldsymbol{x}_{u',r,\rm aug}\boldsymbol{\phi}_{u,r,\rm MF}^H) =
    \\
    \begin{cases}
        M\beta_{ru'}\psi_{ru'}\biggl(\frac{\sin(\pi(m-n)\tau_{uu',r}/\tau_{\rm p})}{\sin(\pi(m-n)/\tau_{\rm p})}\biggl)^2 \hspace*{0.2in}
        \text{$t_{u,r} \leq t_{u',r}$}\\
        M\beta_{ru'}\psi_{ru'}\biggl(\frac{\sin(\pi(m-n)\tau_{uu',r}/\tau_{\rm p})}{\sin(\pi(m-n)/\tau_{\rm p})}\biggl)^2\\
		\hspace*{0.4in}+ M\beta_{ru'}\psi_{ru'}(t_{u'r} - t_{u,r}) \hspace{0.4in} \text{$t_{u,r} > t_{u',r}$}\numberthis
    \end{cases}
\end{gather*}

\subsection{Extended DFT}
Theoretical analysis of the previous two methods reveals that with asynchronous reception, interference from all UEs persists after matched filtering, whereas in the synchronous case, only co-pilot UEs cause interference. Consequently, using these methods results in significantly lower channel estimation accuracy compared to the synchronous scenario.


To cope with this, we propose using DFT sequences with a cyclic extension. Assume that the DFT pilot sequence assigned to the user $u$ is $\boldsymbol{\phi}_u = [\boldsymbol{\Phi}]_m\in\mathbb{C}^{1\times \tau_{\rm p}}$. Then, assume the extended length is $\tau_{\rm ex}$, the pilot sequence being transmitted by the UE $u$ is $\boldsymbol{\phi}_u\in\mathbb{C}^{1\times(\tau_{\rm p} + \tau_{\rm ex})}$ which can be expressed as,
\begin{align}\label{eqn: extended DFT}
	&\boldsymbol{\phi}_{u,\rm DFT_{\rm ex}} = [\boldsymbol{\zeta}_u[0],...,\boldsymbol{\zeta}_u[\tau_{\rm p} - 1],\boldsymbol{\zeta}_u[0],...,\boldsymbol{\zeta}_u[\tau_{\rm ex} - 1]]
	\nonumber\\
	& =
	[\boldsymbol{\zeta}_u[0],...,\boldsymbol{\zeta}_u[\tau_{\rm p} - 1],\boldsymbol{\zeta}_u[\tau_{\rm p}],...,\boldsymbol{\zeta}_u[\tau_{\rm p}+\tau_{\rm ex} - 1]]
\end{align}
where
\begin{align}
\boldsymbol{\zeta}_u[n] = \boldsymbol{\phi}_u[n]
=
e^{j\frac{2\pi m (n)}{\tau_{\rm p}}}
\end{align}
%
%
%
For DFT sequences, it has a special property that the step size between any two adjacent entries is always $e^{j\frac{2\pi m}{\tau_{\rm p}}}$. Therefore, the second equality of~\eqref{eqn: extended DFT} is always true.

\begin{figure}[t] 
	\centering
	\includegraphics[width=0.85\columnwidth]{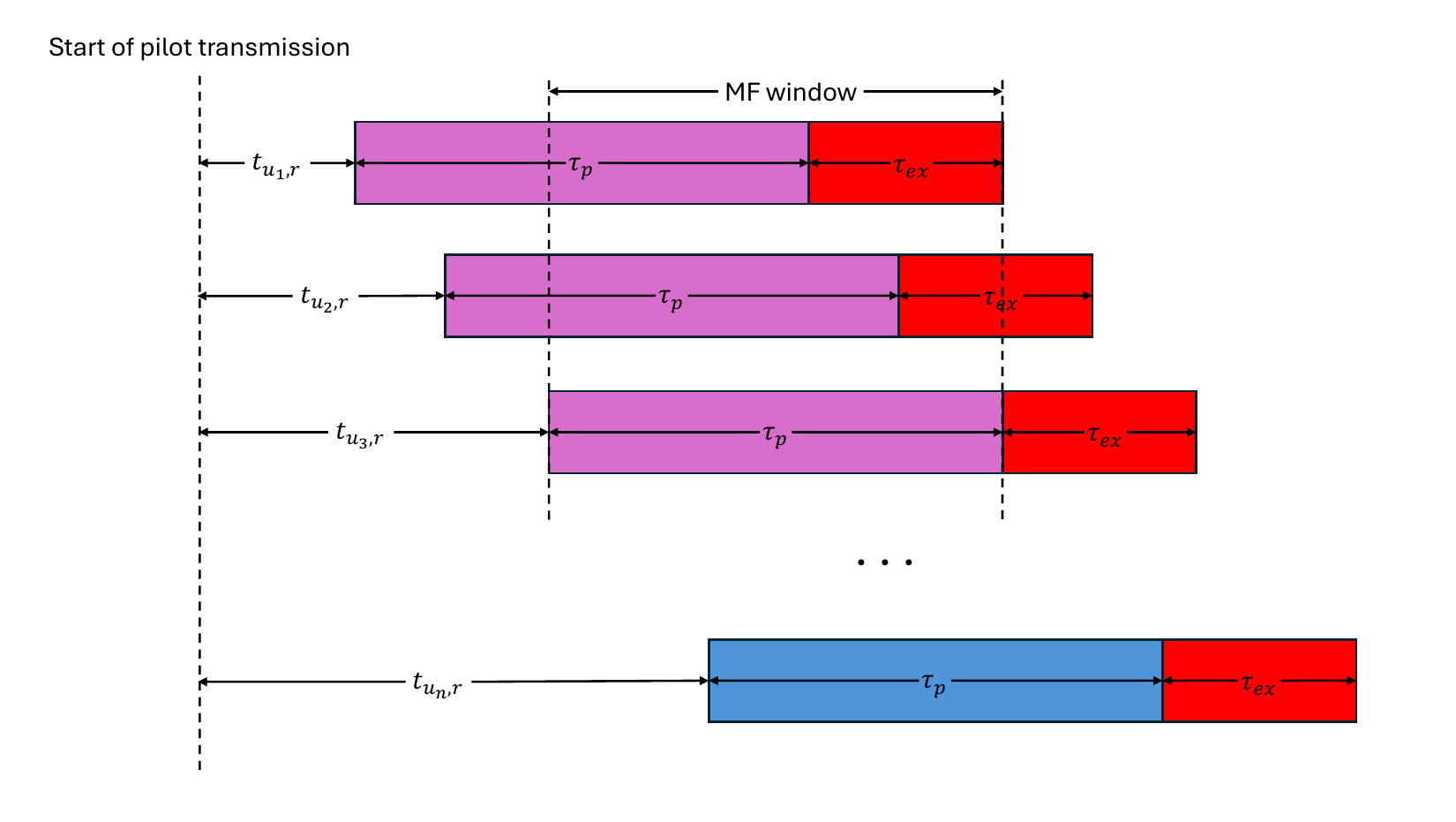}
	\vspace{-1em}
	\caption{Asynchronous Reception with extended DFT}
	\label{fig:Asynchronous Reception with extended DFT}
	\vspace{-2em}
\end{figure}

Now, if the receiver (AP $r$) receives another pilot sequence $\boldsymbol{\phi}_{u'}$, it will first determine an MF window. Assume that $t_{u,r} < t_{u',r}$, then the MF window starts from the arrival of $\boldsymbol{\phi}_{u'}$, and ends at the end of $\boldsymbol{\phi}_{u}$. Besides, the MF window is required to have a length of $\tau_{\rm p}$. Therefore, the received signal model after the MF at the AP $r$ can be written as follows,
\begin{align}
&\boldsymbol{\phi}_{u,r,\rm MF} = \biggl[\boldsymbol{0}^{1\times t_{\rm w,r}},\boldsymbol{\zeta}_u,\boldsymbol{0}^{1\times (t_{\rm max,r} - t_{\rm w,r})}\biggl]\\
   & \boldsymbol{Y}_r\boldsymbol{\phi}_{u,r,\rm MF}^H = \sqrt{p^{\rm ul}} \times
    \nonumber \\
    &
    (\boldsymbol{h}_{ru}\boldsymbol{\phi}_{u,r,\rm aug}\boldsymbol{\phi}_{u,r,\rm MF}^H + \boldsymbol{h}_{ru'}\boldsymbol{\phi}_{u',r,\rm aug}\boldsymbol{\phi}_{u,r,\rm MF}^H)
    + \boldsymbol{Z}_r\boldsymbol{\phi}_{u,r,\rm aug}^H
\end{align}
Here, $t_{\rm w,r}$ is the time delay for the MF window to start at the AP $r$. In the two UEs case, $t_{\rm w,r} = t_{u',r}$, and $t_{\rm max,r} = t_{u',r}$. The $\boldsymbol{\phi}_u\boldsymbol{\phi}_{u,r,\rm aug}^H$ and $\boldsymbol{\phi}_{u'}\boldsymbol{\phi}_{u,r,\rm aug}^H$ can be calculated as,
\begin{align}
    \boldsymbol{\phi}_{u,r,\rm aug}\boldsymbol{\phi}_{u,r,\rm MF}^H &= (e^{j\theta_{u,r}}\boldsymbol{\zeta}_{u})\boldsymbol{\zeta}_u^H
    \nonumber\\
    &= e^{j\theta_{u,r}}\tau_{\rm p}
    \\
    \boldsymbol{\phi}_{u',r,\rm aug}\boldsymbol{\phi}_{u,r,\rm MF}^H &= (e^{j\theta_{u',r}}\boldsymbol{\zeta}_{u'})\boldsymbol{\zeta}_u^H
    \nonumber\\
    &= e^{j\theta_{u',r}}[\boldsymbol{\Phi}]_n[\boldsymbol{\Phi}]_m^H = 0, \ \  m\neq n
\end{align}
Here, $(e^{j\theta_{u,r}}$, $e^{j\theta_{u',r}} )$ represent the phase difference between $(\boldsymbol{\zeta}_u$, $\boldsymbol{\zeta}_{u'})$ and $\left[\boldsymbol{\phi}_{u,r,\rm aug}[t_{\rm w,r}],...,\boldsymbol{\phi}_{u,r,\rm aug}[t_{\rm w,r}+\tau_{\rm p}-1]\right]$ which is the subsegment of $\boldsymbol{\phi}_{u,r,\rm aug}$ that lies within the MF window of the AP $r$. These phase-shifting factors are independent of each other and do not affect the power. \textit{Crucially, the use of the cyclic extension restores the orthogonality between UEs.}

The choice of extension $\tau_{\rm ex}$ is crucial, as it affects performance. While $\tau_{\rm ex}$ must cover interfering UEs, an overly large value reduces throughput. For each AP $r$, we define a set $\mathcal{S}_r$, with $\mathcal{U}_r \subseteq \mathcal{S}_r$ as the set with all significant UEs; these are the UEs close enough to the AP to significantly impact channel estimation performance. Fig.~\ref{fig:Significant User Region} illustrates this, where the red star dot is AP $r$, the pink-filled and blue-filled dots are the UEs being and not being served by AP $r$, respectively. The dashed circle encloses significant UEs, such that it includes all the pink UEs and a few blue UEs. In this case, $t_{\rm w,r} = \rm max_{u\in\mathcal{U}_r}\{t_{u,r}\}$ and $t_{\rm max,r} = \rm max_{u\in\mathcal{U}}\{t_{u,r}\}$. %
Fig.~\ref{fig:Asynchronous Reception with extended DFT} represents the reception time for the users shown in Fig.~\ref{subfig:Significant User Region small}, where the pink and blue sequences are transmitted from the pink and blue UEs in Fig.~\ref{subfig:Significant User Region small}, respectively.

   
\begin{figure}[t]
	\centering
	\includegraphics[width=0.92\columnwidth]{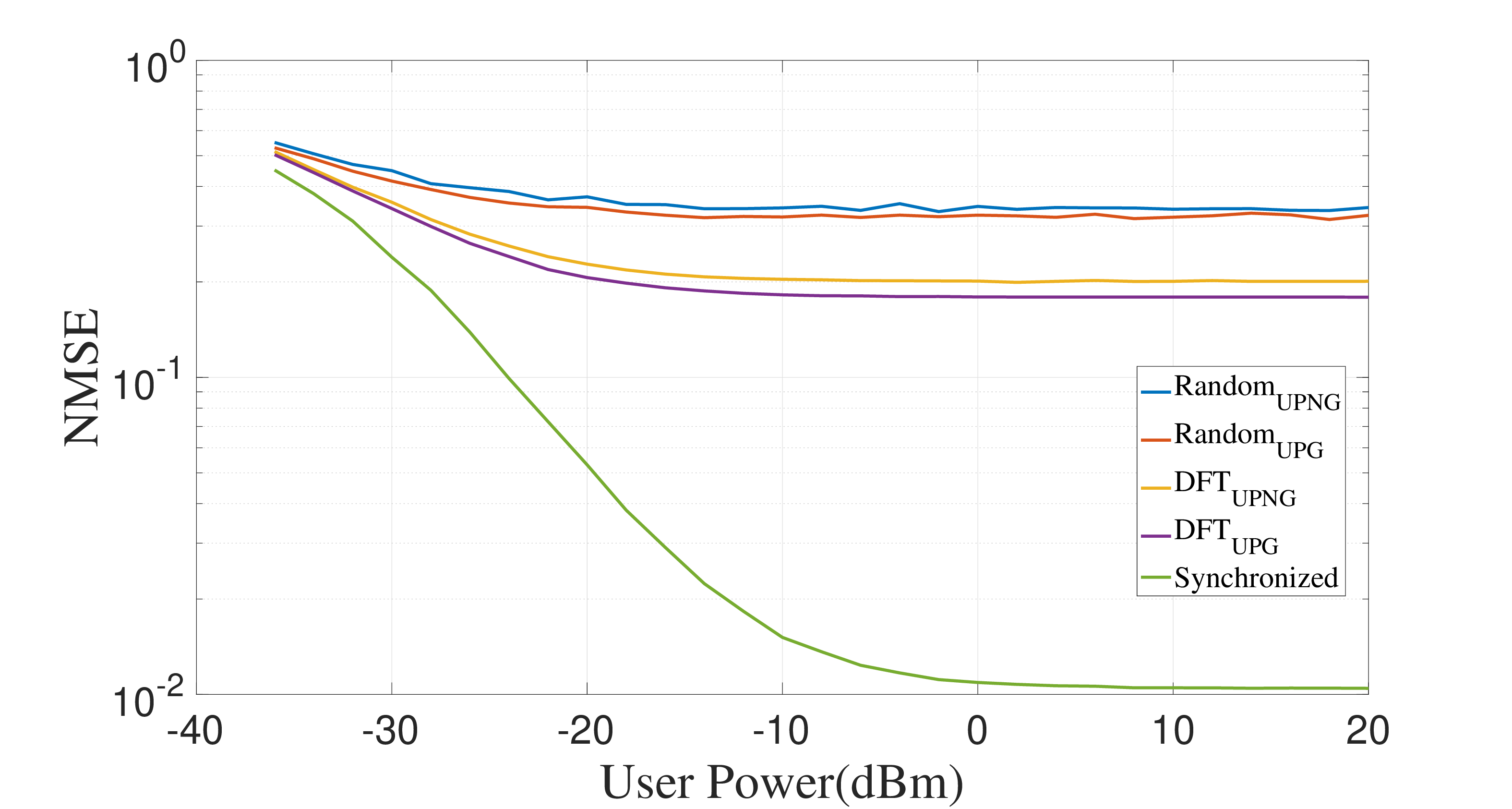}
	\caption{NMSE vs Transmit Power}
	\label{fig:NMSE vs Transmit Power (I)}
	\vspace{-1.4em}
\end{figure}
\begin{figure}[t]
	\centering
	\includegraphics[width=0.92\columnwidth]{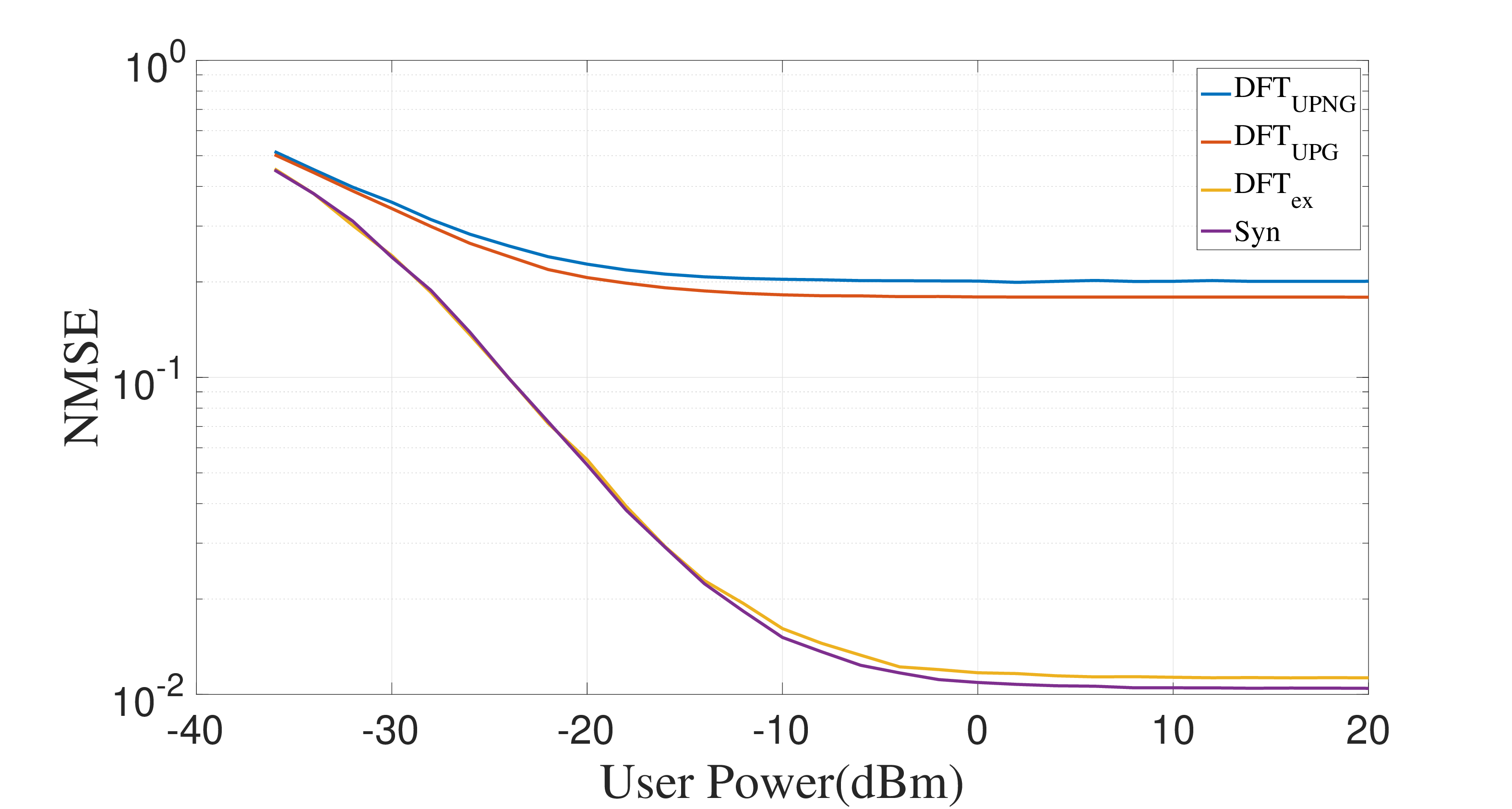}
	\caption{NMSE vs Transmit Power (II)}
	\label{fig:NMSE vs Transmit Power (II)}
	\vspace{-1.8em}
\end{figure}


We choose a reasonable value for the radius of the significant UEs region, and it is defined as,
\begin{align}
    \begin{split}
        \rm r_{\mathcal{S}_r} = \tau_{\rm ex}\tau_{smp}c
    \end{split}
\end{align}
Here, $\tau_{\rm smp}$ and $c$ denote the sampling period and speed of light, respectively. The significant UE region can be treated as a synchronous region for channel estimation. Long DFT pilot extensions improve estimation accuracy by expanding the synchronous region. Then, the signal model using extended DFT sequences after MF and LMMSE can be written as,
\allowdisplaybreaks
\begin{align}
    \begin{split}
        \vspace{-5em}
        \Breve{\boldsymbol{y}}_{ru} &= \tau_{\rm p}\boldsymbol{h}_{ru} + \tau_{\rm p}\sum_{u'\in\mathcal{S}_r\bigcap\mathcal{C}_r}\boldsymbol{h}_{ru}
        \\[-5pt]
        &+ \sum_{u'\not\in\mathcal{S}_r}\boldsymbol{h}_{ru}\boldsymbol{\phi}_{u',r,\rm aug}\boldsymbol{\phi}_{u,r,\rm MF}^H + \frac{\boldsymbol{Z}_r\boldsymbol{\phi}_{u,r,\rm aug}^H}{\sqrt{p^{\rm ul}}}
    \end{split}
    \\
    \begin{split}
        \Sigma_{\Breve{\boldsymbol{y}}_{ru}\Breve{\boldsymbol{y}}_{ru}^H} &= \biggl(\tau_{\rm p}^2\beta_{ru}\psi_{ru} + \sum_{u'\in\mathcal{S}_r\bigcap\mathcal{C}_r}\tau_{\rm p}^2\beta_{ru'}\psi_{ru'}
        \\[-5pt]
        &+ \sum_{u'\not\in\mathcal{S}_r}\beta_{ru'}\psi_{ru'}|\boldsymbol{\phi}_{u',r,\rm aug}\boldsymbol{\phi}_{u,r,\rm MF}^H|^2 + \frac{\sigma^2}{p^{\rm ul}}\biggl)\boldsymbol{I}_M
    \end{split}
\end{align}

\begin{figure}[t] 
	\centering
	\includegraphics[width=0.92\columnwidth]{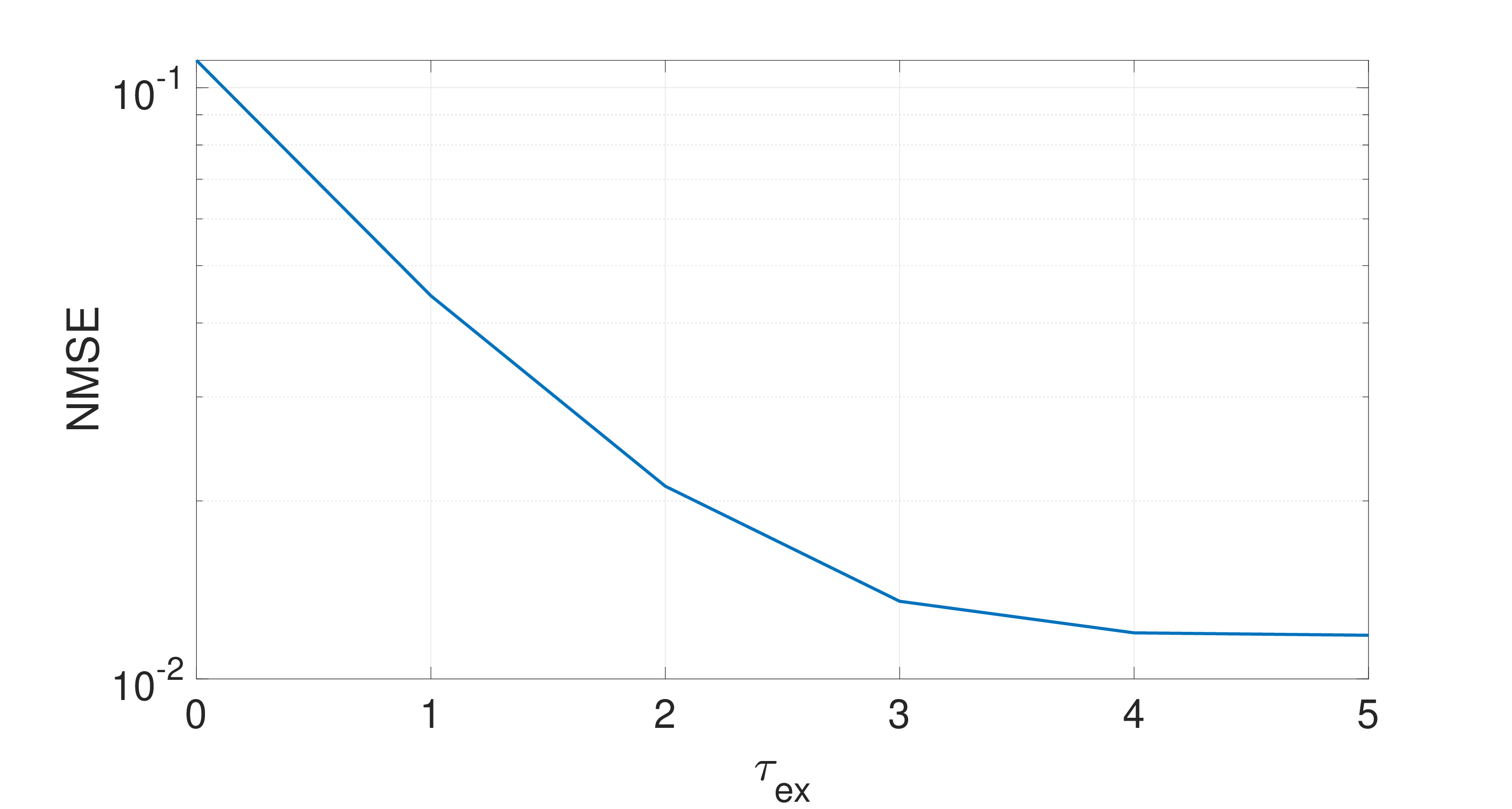}
	\caption{NMSE vs $\tau_{\rm ex}$ with $p^{\rm ul} = 20 \rm dBm$}
	\label{fig:NMSE vs tau_ex}
	\vspace{-1em}
\end{figure}

\begin{figure}[t]
	\centering
	\includegraphics[width=0.92\columnwidth]{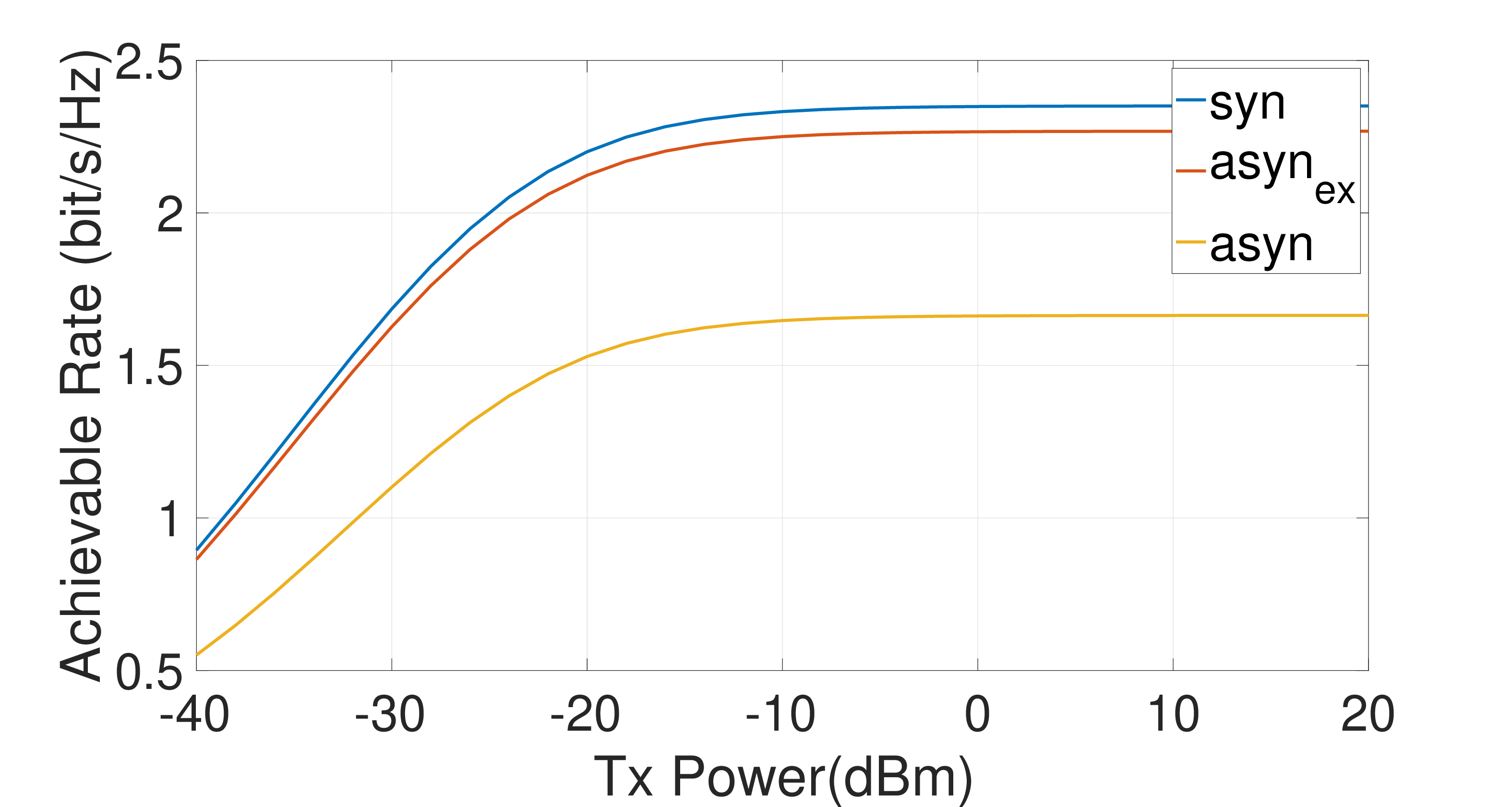}
	\caption{Achievable rate comparison}
	\label{fig:AR_comp}
	\vspace{-1.8em}
\end{figure}

\vspace{-1.5em}
\section{Numerical Results and Analysis}\label{section:results}
%
We simulate an area of 0.7 sq.~km with $|\mathcal{R}| = 70$ APs uniformly distributed. The number of UEs is chosen as a Poisson random variable with a mean of 98. Both APs and UEs are uniformly distributed, i.e., locations of UEs is a Poisson Point Process (PPP). The restricted area radius is $\gamma = 20 \rm m$. The number of UEs served by each AP $r$ is $|\mathcal{U}_r| = 4,\:\forall r$. We use the Walfisch-Ikegami path loss model, $\beta_{ru} = 10^{-11.2427}d_{ru}^{-3.8}$, where $d_{ru}$ denotes the distance between UE $u$ and AP $r$ in km; the shadowing is, $\psi_{ru, \rm dB}\sim \mathcal{N}(0,\sigma^2_{\rm dB}),\:\sigma_{\rm dB} = 4$. The noise power is $\sigma^2 = 1\times 10^{-14}\rm W$. Each AP has $M = 8$ antennas. For simplicity, we assume there are no phase/timing synchronization errors amongst the antennas at each AP. We assume the pilot sequence is transmitted through the entire bandwidth in the channel estimation phase, $\rm BW = 20 \rm MHz$, so $\tau_{\rm smp} = \frac{1}{\rm BW} = 50 \rm ns$. The power, $p^{\rm ul}$, varies from -36 to +20 dBm.

For the extension of the DFT sequence, we set $\tau_{\rm ex} = \tau_{\rm ex, \rm min}$. Here, $\tau_{\rm ex, \min} = \max_{r\in\mathcal{R}}\{\max_{u\in\mathcal{U}_r}\{ t_{u,r}\} - \min_{u'\in\mathcal{U}_r}\{t_{u',r}\}\}$. This choice eliminates interference from all the other UEs being served by the AP $r$, as seen in Fig.~\ref{subfig:Significant User Region small}, in which the significant user region cannot be smaller. We also simulate over the range of 0 $\leq \tau_{\rm ex} \leq 6$.

We first compare the NMSE, as a function of $p^{\rm ul}|_{\rm dBm}$, using random and DFT sequences, in which the pilot length $\tau_{\rm p} = 32$ fixed for two cases, UPG and UPNG with $\rm NMSE = ||\boldsymbol{h}_{ru}-\hat{\boldsymbol{h}}_{ru}||^2/||\boldsymbol{h}_{ru}||^2$. As shown in Fig.~\ref{fig:NMSE vs Transmit Power (I)}, with a short pilot length, DFT sequences outperform random sequences. Adding a guard time (UPG) helps estimate channels more accurately than without a guard time (UPNG). However, even for the best performance among these four cases, $\rm DFT_{UPG}$, the estimation performance is significantly worse than the synchronized case.

Next, we compare the DFT sequences under two cases with extended DFT sequences using $\tau_{\rm ex} = \tau_{\rm ex, \rm min}$; the result is shown in Fig.~\ref{fig:NMSE vs Transmit Power (II)}. Our results show that using the extended DFT sequences improves the estimation performance significantly, essentially equivalent to the synchronous case. Since the extension restores orthogonality to eliminate most of the significant interference within the MF window, it is worth noting that Fig.~\ref{fig:NMSE vs Transmit Power (II)} is the core result of our work. Thanks to the periodicity of the DFT columns, as long as the extension used, $\tau_{\rm ex} \geq \tau_{\rm ex, \min}$, \textit{despite asynchronous reception}, the interference from other UEs being served by the same AP is eliminated. 

As expected, larger $\tau_{\rm ex}$ allows to eliminate more interference and improve channel estimation, as shown in Fig.~\ref{fig:NMSE vs tau_ex}.

To further measure the performance, we calculate the downlink data rate using~\cite[Formulas (19) - (23)]{7827017}, which employs conjugate beamforming. Fig.~\ref{fig:AR_comp} shows the impact of improved channel estimation on achievable rate. The blue curve plots the synchronous case, which acts as the performance upper bound. The yellow curve plots the case when the reception is asynchronous, by using DFT-based pilot sequences without any extension. Clearly, the asynchronous reception causes a significant loss in achievable rate. Finally, the red curve, named as $``\rm asyn_{ex}"$ represents the rate by using the extended DFT pilot sequences. Using the proposed method channel estimation the spectral efficiency increases by about 0.6 bits/s/Hz. Essentially, we have re-established synchronous transmission for the system at the cost of the time and power sacrificed for the extension.


\vspace{-0.5em}
\section{Conclusions}\label{section:conclusion}
\vspace{-0.5em}
In this paper, we investigated asynchronous reception in a cell-free system, while focusing on channel estimation. To address the issue of asynchronous reception, we develop extended DFT sequences using a cyclic shift. The periodic property of DFT sequences enables orthogonality to be restored within the MF window, and hence, synchronous reception at the APs is accomplished within the channel estimation phase. Our simulations confirm the theory developed, showing performance comparable to synchronous reception case.

It is evident that increasing the cyclic extension for the pilot sequences enhances the ability to handle asynchronous reception from UEs across a larger area. However, this also consumes additional time resources. Consequently, striking an optimal balance between time overhead and estimation accuracy is a topic worthy of future exploration.
\appendix\label{section:appendix}
Here is the detailed derivation of~\eqref{DFT cross},
{\allowdisplaybreaks
\begin{align}
    \boldsymbol{x}_{u',r,\rm aug}\boldsymbol{\phi}_{u,r,\rm MF}^H 
    &= \omega^{m(\tau_{\rm p} - \tau_{uu',r})}\Sigma
\end{align}

where
\vspace{-1em}
\begin{align}
    \Sigma &= \sum_{i=0}^{\tau_{uu',r}-1}\omega^{(m-n)i}\nonumber\\
        (\omega^{m-n}-1)\Sigma &= \omega^{(m-n)(\tau_{uu',r})} - 1
        \nonumber\\
        \Sigma &= (\omega^{(m-n)\tau_{uu',r}} - 1)/(\omega^{m-n}-1)
\end{align}}

To find~\eqref{DFT cross power}, we find the absolute value squared of~\eqref{DFT cross},
{\allowdisplaybreaks
\begin{align}
    &||\boldsymbol{x}_{u',r,\rm aug}\boldsymbol{\phi}_{u,r,\rm MF}^H||^2
    \nonumber \\
    & 
    = \biggl(\frac{\omega^{(m-n)(\tau_{uu',r})} - 1}{\omega^{m-n}-1}\biggl)^*\biggl(\frac{\omega^{(m-n)(\tau_{uu',r})} - 1}{\omega^{m-n}-1}\biggl)
    \nonumber\\
    & 
    = \frac{1-2\Re\left\{\omega^{(m-n)\tau_{uu',r}}\right\}+1}{1-2\Re\left\{\omega^{m-n}\right\}+1}\nonumber\\
    & 
    = \frac{1-\cos(2\pi(m-n)\tau_{uu',r}/\tau_{\rm p})}{1-\cos(2\pi(m-n)/\tau_{\rm p})} = \frac{\sin^2(\pi(m-n)\tau_{uu',r}/\tau_{\rm p})}{\sin^2(\pi(m-n)/\tau_{\rm p})}
\end{align}}

\vspace{-1em}

\bibliography{biblio}

\begin{thebibliography}{10}

\bibitem{9349624}
W.~Jiang, B.~Han, M.~A. Habibi, and H.~D. Schotten, ``The road towards 6g: A
  comprehensive survey,'' {\em IEEE Open Journal of the Communications
  Society}, vol.~2, pp.~334--366, 2021.

\bibitem{9113273}
J.~Zhang, E.~Björnson, M.~Matthaiou, D.~W.~K. Ng, H.~Yang, and D.~J. Love,
  ``Prospective multiple antenna technologies for beyond 5g,'' {\em IEEE JSAC},
  vol.~38, no.~8, pp.~1637--1660, 2020.

\bibitem{9650567}
H.~A. Ammar, R.~Adve, S.~Shahbazpanahi, G.~Boudreau, and K.~V. Srinivas,
  ``User-centric cell-free massive mimo networks: A survey of opportunities,
  challenges and solutions,'' {\em IEEE Communications Surveys \& Tutorials},
  vol.~24, no.~1, pp.~611--652, 2022.

\bibitem{8845768}
E.~Björnson and L.~Sanguinetti, ``Making cell-free massive {MIMO} competitive
  with {MMSE} processing and centralized implementation,'' {\em IEEE TWC},
  vol.~19, no.~1, pp.~77--90, 2020.

\bibitem{8969384}
H.~A. Ammar and R.~Adve, ``Power delay profile in coordinated distributed
  networks: User-centric v/s disjoint clustering,'' in {\em 2019 IEEE
  GlobalSIP}, pp.~1--5, 2019.

\bibitem{Nasir_2016}
A.~A. Nasir, S.~Durrani, H.~Mehrpouyan, S.~D. Blostein, and R.~A. Kennedy,
  ``Timing and carrier synchronization in wireless communication systems: a
  survey and classification of research in the last 5 years,'' {\em EURASIP
  Journal on WCN}, vol.~2016, Aug. 2016.

\bibitem{8676341}
H.~Yan and I.-T. Lu, ``Asynchronous reception effects on distributed massive
  mimo-ofdm system,'' {\em IEEE Transactions on Communications}, vol.~67,
  no.~7, pp.~4782--4794, 2019.

\bibitem{10459246}
U.~Kunnath~Ganesan, T.~T. Vu, and E.~G. Larsson, ``Cell-free massive mimo with
  multi-antenna users and phase misalignments: A novel partially coherent
  transmission framework,'' {\em IEEE Open Journal of the Communications
  Society}, vol.~5, pp.~1639--1655, 2024.

\bibitem{ganesan2023beamsyncovertheairsynchronizationdistributed}
U.~K. Ganesan, R.~Sarvendranath, and E.~G. Larsson, ``Beamsync: Over-the-air
  synchronization for distributed massive mimo systems,'' 2023.

\bibitem{9531352}
J.~Li, M.~Liu, P.~Zhu, D.~Wang, and X.~You, ``Impacts of asynchronous reception
  on cell-free distributed massive mimo systems,'' {\em IEEE Trans. on
  Vehicular Technology}, vol.~70, no.~10, pp.~11106--11110, 2021.

\bibitem{10517887}
Z.~Li and R.~Adve, ``Uplink resource allocation optimization for user-centric
  cell-free mimo networks,'' {\em IEEE TWC}, pp.~1--1, 2024.

\bibitem{8277240}
E.~Björnson, J.~Hoydis, and L.~Sanguinetti, {\em Massive MIMO Networks:
  Spectral, Energy, and Hardware Efficiency}.
\newblock 2017.

\bibitem{7827017}
H.~Q. Ngo, A.~Ashikhmin, H.~Yang, E.~G. Larsson, and T.~L. Marzetta,
  ``Cell-free massive mimo versus small cells,'' {\em IEEE TWC}, vol.~16,
  no.~3, pp.~1834--1850, 2017.

\end{thebibliography}
\bibliographystyle{ieeetr}
\end{document}